\documentclass[sn-mathphys]{sn-jnl}
\jyear{2021}%

\theoremstyle{thmstyleone}%

\theoremstyle{thmstyletwo}%
\newtheorem{remark}{Remark}%

\usepackage{subcaption}
\usepackage{bbm}
\usepackage{amsmath}

\theoremstyle{thmstylethree}%
\newtheorem{definition}{Definition}%

\title[]{Quantifying Decarbonization Speed Across Climate Scenarios}
\date{\Today}
\author*[]{\fnm{Fangyuan} \sur{ Zhang}}\email{fangyuan.zhang@edhec.edu}

\affil[]{\orgdiv{EDHEC Climate Institute}, \orgname{EDHEC Business School}, \orgaddress{\street{393 Promenade des Anglais}, \city{Nice}, \postcode{06200}, \country{France}}}

\abstract{\textcolor{black}{In this work, we analyze 126 publicly available IAM climate scenarios modeled by six leading teams in climate science. We define a simple numerical metric that measures the decarbonization speed implied by each IAM scenario. With this metric, the narrative based, high-dimensional time series scenario datasets can be ranked and compared in a transparent way. We find that the ranking of IAM scenarios according to the decarbonization speed is consistent with their representative concentration pathway assumptions, showing that the decarbonization metric is a useful summary of a scenario's mitigation policy. We further construct an empirical distribution and a fitted parametric distribution of the decarbonization speed estimates. Key statistics such as mean, median and their confidence intervals by the bootstrap resample technique are also reported.}  }

\begin{document}
\maketitle
\renewcommand{\thefootnote}{}
\renewcommand{\thefootnote}{\arabic{footnote}}



\maketitle
\newpage
\section{Introduction}\label{sec_intro}
Climate scenarios are a fundamental tool for studying the interplay of economic activity, transition policy and climate change. \textcolor{black}{The construction of a climate scenario usually starts by narratives on their socioeconomic development and mitigation objectives. Then, these narratives are quantified by the Integrated Assessment Models (IAM). In the end, the climate scenarios are provided as high-dimensional time series projections of variables relevant for transition and mitigation. In reality, these IAM scenarios further form a basis for the framework in IPCC assessments and CMIP5/6 climate model experiments.}

The IAM scenario datasets cover a broad range of variables relevant for transition, from macroeconomic variables such as GDP and consumption, to climate variables such as CO$_2$ concentration and radiative forcing. However, the high dimensionality of the scenario datasets makes it difficult to directly compare these scenarios. In practical applications, always the analysis is restricted to a subset of a few scenarios. An overview of the full ensemble of IAM scenarios is lacking. 


This work provides one potential approach to address this challenge. We define a simple metric that provides a numerical summary of the decarbonization speed implied by each IAM scenario. The decarbonization speed provides a smooth measure of the improvement of carbon intensity trajectory, which is at the center of a mitigation pathway. Through this metric, the narrative based, high dimensional scenario datasets can be ranked and compared on a common scale. Table \ref{tab:overview_ambition} classifies 126 IAM scenarios according to their decarbonization speed. We find that the ranking of the scenarios are consistent with their mitigation objective assumptions. This confirms that the decarbonization speed is a useful summary of the implied mitigation schedule in each scenario. 

The decarbonization speed also has an intuitive interpretation, i.e., it enables to compute the time to halve the initial carbon intensity. For example, the carbon intensity at 2010 is around 0.24 KgCO$_2$/kwh.\footnote{A formal definition of carbon intensity using scenario variables is given in Section \ref{subsec_carbon intensity}.}Around this time, more than 85\% of primary energy are supplied by fossil fuels. The carbon intensity trajectory directly reflects an economy's dependence on fossil fuels. Using the decarbonization speed, we find that 90\% of IAM scenarios are projected to halve its initial level by 2100. The historical data shows that (Fig. \ref{fig:CarbonIntensityHistory}) European Union, the fasted decarbonizing area, spent almost sixty years to halve its carbon intensity at 1965. However, around 70\% of IAM scenarios are projected to halve the initial carbon intensity faster than EU, which may suggest that the IAM scenarios have an optimistic overview in terms of decarbonization schedule.

We analyze all 126 publicly available IAM scenarios. Using the 126 decarbonization speed estimates, we construct an empirical distribution of the scenarios. To represent these scenarios even more concisely, we fit a parametric distribution to the empirical distribution (Fig. \ref{fig:lognormaldensity}). Comparison of some key statistics such as mean and median of the estimates shows that the fitted lognormal distribution captures well the main features of the original estimates. This suggests that such a stylized parametrization can be regarded as a concise representation of the IAM scenarios in terms of the decarbonization schedule, which may facilitate further quantitative analysis of the IAM scenarios. However, there are also limitations of this approach. In Section \ref{subsec_limits}, we discuss several caveats and limits of the current approach.

This work is related to probabilistic and uncertainty analysis of climate change.
The analysis of climate change involves several layers of deep uncertainty. Three major sources of uncertainty are identified: 1) uncertainty arising in the response of climate variables, e.g., temperature change and precipitation,  to a given emissions path; 2) uncertainty of future emissions scenarios, which depends on the transition policy, economic growth and technology development; 3) uncertainty in the impact of climate change on the human society including economic productivity, agriculture and biodiversity (see for example \cite{knutti2008review} and \cite{dessai2004does}). These uncertainties are difficult to analyze using the historical data alone. Because of the incomplete knowledge, in many cases, it is also difficult to explore the uncertainty via simulation. A variety of methods have been developed to address these challenges. 

One important source of these uncertainties stems from the lack of complete knowledge, in contrast to random variability (i.e., noise). This motivates the use of a Bayesian and subjective probabilistic analysis for these uncertainties (\cite{schneider1999uncertainties}, \cite{schneider2001dangerous}, \cite{pittock2001probabilities}). Bayesian probabilistic analysis enables constant updating of the estimate as new information is available. For example, a Bayesian Monte Carlo framework is used to study the distribution of the climate variables (such as temperature and precipitation) given the prior distributions for the model parameters (\cite{new2000representing}). Expert elicitation is also a common way to conduct subjective probabilistic analysis. It extracts the judgements from domain experts about uncertain quantities and formalizes the corresponding subjective probability distribution. \textcolor{black}{For example, collecting the estimates of social-cost-of-carbon from different generations of economists formulates the economists' distribution of theoretical carbon price (\cite{tol2023social}). Expert elicitation can also be applied to quantify the uncertainty in climate sensitivity and estimate the likelihood of scenarios (\cite{morgan1995subjective},\cite{venmans2022unconditional},\cite{rebonato2025climate}, and \cite{kainth2025developing}).} 

In addition to subjective Bayesian approaches, some studies apply an imprecise probability technique using belief functions to investigate the uncertainty in the climate outcomes (\cite{kriegler2005utilizing}, \cite{luo1997using}, \cite{hall2007imprecise}). The use of Multimodel Ensemble is also a common way to study the uncertainty in climate change. Projections from multiple climate models are combined to characterize model uncertainty (\cite{raisanen2001probability}). Other studies that provide a probabilistic analysis of the climate uncertainties can be found for instance in \cite{webster2002uncertainty}, \cite{jones2000analysing} and \cite{lempert2000robust}. For a review of the relevant study on the uncertainty in climate change, we recommend \cite{knutti2008review}, \cite{katz2002techniques}, \cite{katzav2021appropriate}, \cite{yohe1999risk}, \cite{dessai2004does} and the references therein. 

In addressing the deep uncertainty inherent in climate change, climate scenarios have been proposed as one important tool (\cite{werner1997proposal}, \cite{mote2011guidelines}). However, scenarios are usually provided without associated likelihoods. Therefore, existing studies using climate scenarios mainly focus on a subset of scenarios of particular interest or treat all scenarios equally (\cite{mastrucci2025towards}, \cite{kay2023using}). \textcolor{black}{This work provides a potential framework to give a probabilistic analysis of the IAM climate scenarios. Through a parsimonious metric and its empirical distribution, we provide a structured representation of the available IAM scenario ensemble . Unlike probabilistic climate projections derived from ensemble modeling or subjective probability assessments, our approach summarizes the mitigation ambition embedded in published scenario pathways without introducing additional prior assumptions.}

The rest of the paper is organized as follows. The second section gives the definition and interpretation of the decarbonization speed. In particular, Table \ref{tab:overview_ambition} provides an overview of the 126 decarbonization estimates. The third section constructs an empirical distribution and a fitted lognormal distribution of the estimates. Key statistics such as mean and quantiles of the decarbonization estimates are reported together with the confidence interval by the bootstrap method. Moreover, Section \ref{subsec_limits} discusses the caveats and limitations of the current approach. The last section concludes. 
\section{Decarbonization Speed: Definition and Interpretation}\label{sec_metric}
This section introduces the definition of the decarbonization speed and its interpretation in mitigation ambition. We first describe the dataset used in our analysis.
\subsection{Description of the dataset}\label{subsec_data}
\textcolor{black}{The datasets analyzed in this study consist of Integrated Assessment Model (IAM) scenarios \citep{Riahi2017} maintained by the International Institute for Applied Systems Analysis (IIASA).\footnote{The link to the database: \hyperlink{https://ssp.legacy.ece.iiasa.ac.at/legacy-sspdb/dsd?Action=htmlpage&page=10}{IIASA}.} These scenarios form the basis for the scenario framework used in IPCC assessments and for the forcing pathways employed in CMIP6 climate model experiments, underscoring their relevance for both climate and policy analysis.\footnote{IPCC: The Intergovernmental Panel on Climate Change is a leading United Nations body responsible for assessing the scientific evidence on climate change. CMIP6: The Coupled Model Intercomparison Project Phase 6 is an international framework coordinating climate model experiments under selected IAM scenarios.}}

\textcolor{black}{These scenarios are defined by a combination of economic and policy narratives. The economic narratives are represented by the Shared Socioeconomic Pathways (SSPs), which specify key socioeconomic drivers such as population growth, economic development, and technological progress. For details on the five SSPs, we recommend \cite{Vuuren2017,Fricko2017,Fujimori2017,Kriegler2017,Calvin2017}. The policy narratives correspond to alternative representative concentration (RCP) by the end of the century, reflecting different levels of climate mitigation and associated warming outcomes \citep{Rogelj2018, gmd-12-1443-2019}. The name of a scenario, e.g., SSP1-19 indicates its socioeconomic assumption, SSP1, and its policy target, RCP19. RCP19 means the radiative forcing by 2100 is around 1.9 W/m$^2$ in this scenario. There are in total five SSPs, SSP1 to SSP5, and six RCPs, RCP19, 26, 34, 45, 60 and Baseline. Baseline does not impose an explicit forcing target. Instead, the resulting forcing level emerges endogenously from the underlying SSP assumptions. Typical forcing levels by 2100 are around 5~W/m$^2$ for SSP1-Baseline, around 6.5~W/m$^2$ for SSP2- and SSP4-Baseline, around 7~W/m$^2$ for SSP3-Baseline, and around 8.5~W/m$^2$ for SSP5-Baseline.}

\textcolor{black}{These scenario narratives are quantified by several leading climate research institutes using Integrated Assessment Models. Table \ref{tab:IAM} gives information of these modeling teams and their IAMs. Each modeling team independently provides datasets for a range of SSP-RCP scenarios. Note that the scenarios by MESSAGE-GLOBIOM do not cover SSP4 and SSP5. The scenarios by REMIND-MAGPIE do not cover SSP3 and SSP4. The remaining four IAMs cover all SSPs and RCPs.}

\textcolor{black}{Each IAM scenario provides a comprehensive dataset describing trajectories across multiple key dimensions of the transition such as energy use pathways and emissions pathways. These trajectories can be interpreted as potential mitigation pathways to achieve a given radiative forcing target (RCP) under a specified socioeconomic condition (SSP). Each IAM scenario dataset contains eight categories of variables, and within each category, IAM variables are classified in greater details. For example, the energy category includes primary energy, secondary energy, energy service, etc. The emissions category includes CO$_2$ emissions and non-CO$_2$ emissions, etc. Table \ref{tab:IAMcategory} provides an overview of the categories and some examples of variables within each category. In addition, each variable is given as a time series projection from 2005 to 2100.}

\textcolor{black}{We analyze all publicly available 126 IAM scenarios produced by six modeling teams (corresponding to six distinct IAM frameworks). Although all scenarios report a common set of core variables, they should be interpreted as an ensemble of alternative but internally consistent transition pathways.}

\begin{table}[!htbp]
    \centering
    \begin{tabular}{ccc}
    \hline
    \hline
     IAM & Institute &   \# Scenarios  \\
     \hline
      AIM/CGE & NIES & 24\\
      GCAM & PNNL & 23 \\
      IMAGE & PBL& 23\\
      MESSAGE-GLOBIOM   &IIASA&15 \\
      REMIND-MAGPIE & PIK& 17 \\
      WITCH-GLOBIOM & FEEM& 24\\
      \hline
         \hline
    \end{tabular}
    \caption{Six modeling teams and their IAMs.}
    \label{tab:IAM}
\end{table}

\begin{table}[!htbp]
    \centering
    \begin{tabular}{rl|l}
    \hline
    \hline
     &  Category  & Example Variables \\
       \hline
    1& GDP and Population    & GDP (ppp); Population \\
     2&   Energy& Primary Energy; Energy Service \\
      3& Land Cover & Cropland; Forest\\
      4& Emissions & CO$_2$; Kyoto Gases\\
      5& Climate & CO$_2$ concentration; CO$_2$ forcing \\
      6 &Agricultural Indicators & Crops demand; Livestock production\\
      7& Economic Indicators & Consumption; Carbon Price \\
      8 & Technological Indicators & Electricity Capacity; Solar Capacity\\
       \hline
         \hline
    \end{tabular}
    \caption{A structural overview of the scenario datasets.}
    \label{tab:IAMcategory}
\end{table}
\subsection{Carbon Intensity}\label{subsec_carbon intensity}

\begin{figure}[!htbp]
    \centering
    \includegraphics[width=0.75\linewidth]{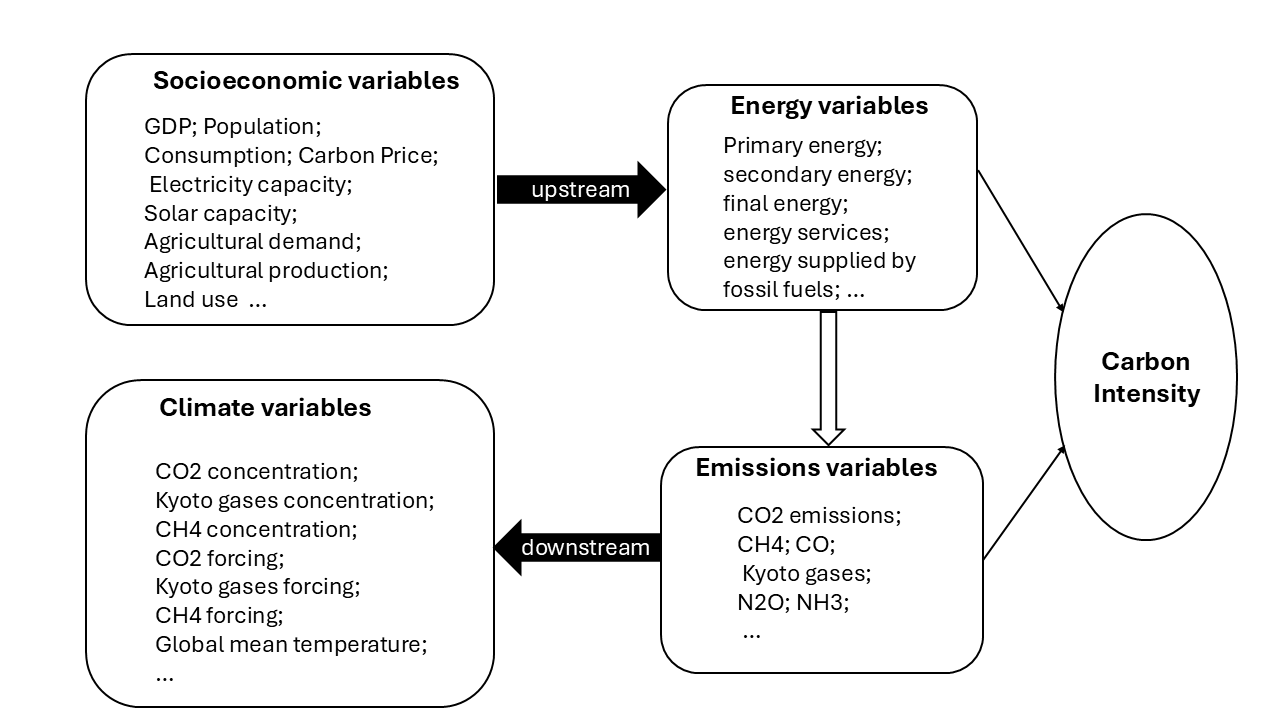}
    \caption{A schematic of the reported variables by IAM datasets.}
    \label{fig:schematicIAM}
\end{figure}

\textcolor{black}{While IAM datasets are rich in reported variables, the key dynamics relevant for mitigation are primarily governed by energy-system choices and their emissions consequences. This observation allows us to classify variables outside the energy and emissions blocks into two broad groups. The first group consists of variables upstream of the energy system. For example, GDP and population belong to this group. Their choices closely determine the scale and structure of energy demand. The second group consists of climate variables that arise as downstream outcomes of emissions pathways. For instance, carbon concentration and radiative forcing belong to this group. Figure \ref{fig:schematicIAM} provides a schematic representation of this classification.} Considering the fact that CO$_2$ emissions are the major contributor to anthropogenic global warming, Carbon Intensity (soon detailed in Definition \ref{def:ci}), i.e., CO$_2$ emissions per unit of energy production, works as a comprehensive and interpretable summary statistic of an IAM scenario's mitigation pathway. 

\begin{figure}[!htbp]
    \centering
    \includegraphics[width=0.75\linewidth]{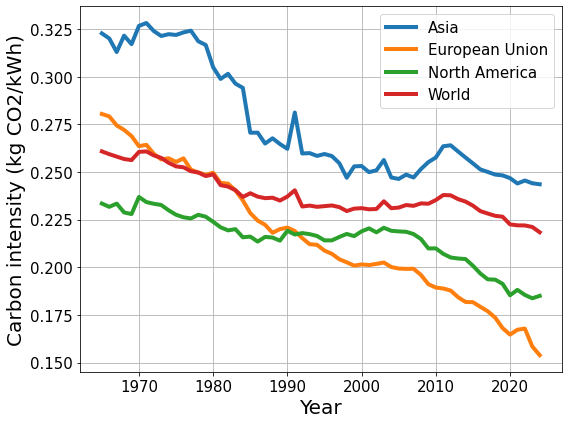}
    \caption{Historical data on carbon intensity by regions. Data source: Global Carbon Budget (2025); U.S. Energy Information Administration (2025); Energy Institute-Statistical Review of World Energy (2025) - with major processing by Our World in Data.}
    \label{fig:CarbonIntensityHistory}
\end{figure}

The trajectory of carbon intensity encapsulates information on several fundamental aspects of the transition. Approximately 80–85\% of global CO$_2$ emissions originate from fossil fuel combustion and industrial processes,\footnote{The remaining emissions mainly arise from land-use change and forestry.} while more than 80\% of global primary energy production is supplied by fossil fuels. Fossil fuel combustion is therefore the dominant contributor to anthropogenic CO$_2$ emissions and, consequently, to global warming. The evolution of carbon intensity directly reflects an economy’s dependence on fossil fuels. Figure~\ref{fig:CarbonIntensityHistory} illustrates the historical trajectories of carbon intensity across different regions. Overall, carbon intensity has declined over time. This downward trend can be attributed primarily to improvements in energy efficiency and the increasing deployment of low-carbon energy use (such as neuclear) and renewable energy sources.

\begin{figure}[!htbp]
    \centering
    \includegraphics[width=0.75\linewidth]{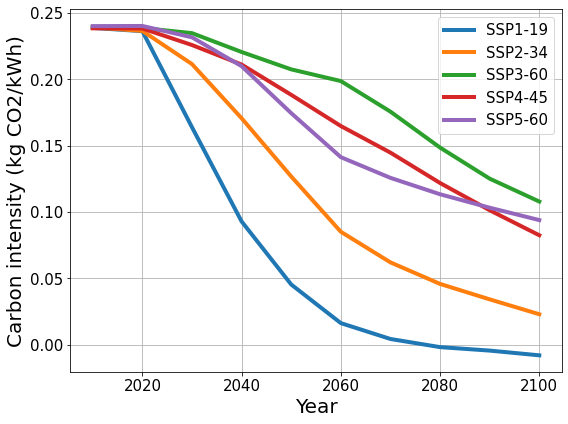}
    \caption{Carbon Intensity from five selected IAM scenarios. The IAM is AIM/CGE.}
    \label{fig:carbonintensitySSP}
\end{figure}

In forward-looking IAM scenarios, carbon intensity is also a key indicator of the level of transition across scenarios. Since it is not a variable directly reported by IAM scenarios, we give a formal definition of carbon intensity using IAM variables.
\begin{definition}{Carbon Intensity}\label{def:ci}\\
Carbon Intensity is defined as the ratio of total CO2 emissions resulting from the production and use of all fossil fuel-based energy sources and industrial processes to total primary energy consumption. I.e.,
\begin{equation}\label{eq:carbonintensity}
    \text{Carbon Intensity}:=\frac{\text{CO$_2$(Fossil fuels and Industry)}}{\text{Total Primary Energy}}.
\end{equation}
\end{definition}
When computing the carbon intensity in IAM scenarios, we exclude the CO$_2$ emissions from land use change in order to focus on the contribution of energy transition, which plays a central role in mitigation pathways. Figure \ref{fig:carbonintensitySSP} illustrates carbon intensity trajectories from five selected IAM scenarios. As mentioned above, the name of an IAM scenario indicates its perticular SSP and RCP assumptions. A higher RCP level indicates a warmer scenario. Carbon intensity trajectories are projected to decline over time, consistent with historical trends (Figure \ref{fig:CarbonIntensityHistory}). Scenarios associated with lower forcing targets (cooler scenarios) exhibit substantially faster declines in carbon intensity, reflecting their more ambitious mitigation narratives. Although SSP3-60 and SSP5-60 reach the same final forcing level, they display distinct carbon intensity pathways. These differences arise from contrasting socioeconomic assumptions that influence energy efficiency improvements. Carbon intensity therefore provides a comprehensive indicator summarizing the overall mitigation ambition embedded in each scenario.

\subsection{Decarbonization speed}\label{subsec_decarbon}
The carbon intensity trajectory can be computed within each IAM scenario. We then introduce a one-dimensional metric that quantifies the speed of improvement in these trajectories. The resulting decarbonization speed provides a simple numeric summary of the overall decarbonization trend. It also admits a clear interpretation in terms of the timing of the transition. Although the IAM scenarios are constructed under complex and multidimensional assumptions, the decarbonization speed metric enables straightforward comparison and ranking of scenarios on a common scale. We begin by defining the decarbonization rate for an individual scenario. 

\begin{figure}[!htbp]
    \centering
    \includegraphics[width=0.75\linewidth]{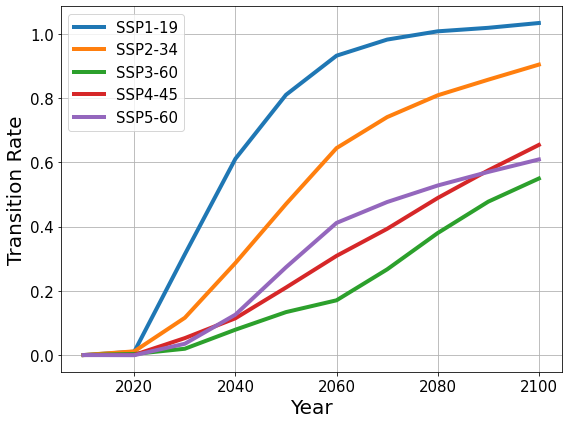}
    \caption{The transition rate trajectories in five selected IAM scenarios. The IAM is AIM/CGE.}
    \label{fig:transitionrate}
\end{figure}

\begin{definition}{Decarbonization Rate}\\
    Let $\sigma_i(t0)$ be the initial carbon intensity and $\sigma_i(t)$ be the carbon intensity at time $t$ in scenario $i$, respectively. The decarbonization rate $u_i(t)$ at time $t$ is given as follows:
    \begin{equation}\label{eq:transition}
        u_i(t)= 1-\frac{\sigma_i(t)}{\sigma_i(t0)}.
    \end{equation}
 \end{definition} 
The decarbonization rate defined in Eq.~\eqref{eq:transition} is a relative metric that quantifies the extent of transition in a scenario with respect to a fixed starting point.\footnote{This work exclusively focuses on energy transition. Therefore, transition in this work is equivalent to decarbonizaition. In other contexts, transition may have a broader meaning.} By construction, the decarbonization rate equals zero at the initial time $t0$. As time progresses and carbon intensity declines, the transition rate increases monotonically. An increasing decarbonization rate therefore indicates a reduction in carbon intensity and corresponds to a more pronounced mitigation pathway. Figure~\ref{fig:transitionrate} illustrates the decarbonization rate trajectories computed for five selected IAM scenarios. In the most ambitious mitigation scenario, SSP1-19, the transition rate exceeds one by 2100. This reflects the projection of net negative CO$_2$ emissions toward the end of the century, and hence negative carbon intensity. Such negative emissions are associated with the large-scale deployment of carbon removal and carbon capture and storage technologies along the pathway. In contrast, under SSP3-60 the transition rate remains below 60\% by 2100, consistent with its comparatively high carbon intensity. These contrasting trajectories highlight that the decarbonization rate provides a simple and transparent measure of the extent of carbon intensity reduction achieved under different mitigation pathways.

\begin{definition}{Decarbonization speed}\\
    To measure the speed of decarbonization, we assume that the transition rate follows a parametric function defined below:
\begin{equation}\label{eq:paratransition}
        u_i(t) = u_{\max} (1-\exp(-\theta_i\cdot t)).
    \end{equation}
    where $u_{\max}\geq 1$ denotes the maximum decarbonization rate of all scenarios and $\theta_i>0$ measures the unique decarbonization speed in scenario $i$.
\end{definition}

    \begin{remark}{Below are some remarks on the decarbonization speed.}\\
\begin{enumerate}
    \item Each IAM scenario has a unique decarbonization speed $\theta_i$. The decarbonization rate $u_i(t)$ increases from zero to its maximum value $u_{\max}$; 
    \item At the early stage of the transition, when $u_i(t)$ is close to zero, the carbon intensity stays near its initial level. As $u_i(t)$ increases, the carbon intensity declines and eventually approaches zero. 
    \item If $u_{\max}>1$, the carbon intensity can become negative at some point. This occurs due to the deployment of negative emissions technology, such as biomass energy sources with carbon capture and sequestration (BECCS), which results in net negative emissions. 
    \item The metric $\theta_i$ is a smooth measure of the overall decarbonization speed of a scenario. As IAM datasets are produced with a ten year step size, the smooth speed metric captures well the decarbonization trend in most scenarios. However, the smooth decarbonization metric cannot capture acute changes in trajectores.\footnote{For instance, in the Delayed Transition scenario by NGFS (Network for Greening the Financial System), the transition efforts are postponed to 2030. This detailed feature cannot be explicitly reflected by the decarbonization speed.}  
\end{enumerate}
\end{remark}
\subsection{Estimating the scenario-specific decarbonization speed}\label{subsec_estimate}
   Suppose all scenarios share the same maximum possible decarbonization rate but differ in the speed at which they pursue it. In this case, the scenario-specific decarbonization speed, denoted by $\theta_i$ serves as a quantitative metric for each scenario's transition ambition.

\begin{figure}[!htbp]
       \centering
\includegraphics[width=0.75\linewidth]{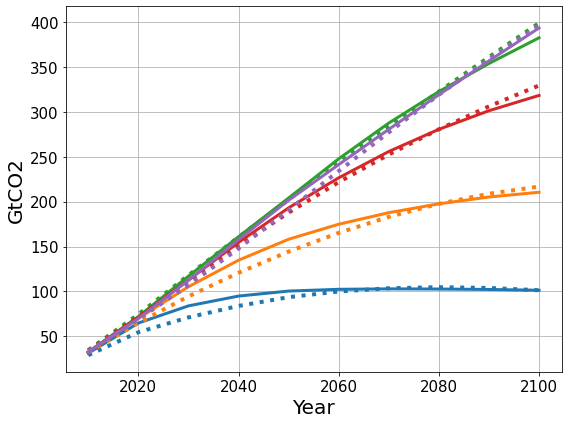}
       \caption{The solid line represents the cumulative emissions in the original datasets and the dotted line represents the reconstructed cumulative emissions with the fitted decarbonization speed.}
       \label{fig:FitCumEmissions}
   \end{figure}

   To estimate the scenario-specific decarbonization speed $\theta_i$, we first compute the decarbonization rate trajectory for each scenario using Eq.~\eqref{eq:transition}. We then identify the maximum decarbonization rate, $u_{\max}$, across all 126 IAM scenarios.\footnote{In our calculation, $u_{\max}\approx 1.52$. It occurs in the scenario SSP4-26 modeled by GCAM. } The value $u_{\max}$ provides an empirical upper bound for the decarbonization rate in Eq. \eqref{eq:paratransition}. In the final step, the decarbonization speed $\theta_i$ of each scenario is obtained through numerical optimization. There are different ways to estimate $\theta_i$ and the accuracy depends on the chosen numerical optimization criterion. In this study, we select the scenario-specific $\theta_i$ that minimizes the distance between cumulative emissions derived from the original IAM scenario and those reconstructed using Eq.~\eqref{eq:paratransition}. This choice is motivated by the fact that cumulative emissions are closely linked to radiative forcing levels and hence to RCP targets. By minimizing the discrepancy in cumulative emissions, we ensure that the reconstructed trajectories remain as consistent as possible with the original IAM scenario narratives. Figure \ref{fig:FitCumEmissions} demonstrates the cumulative emissions in the original datasets and the reconstructed cumulative emissions by the fitted decarbonization speeds of the five selected IAM scenarios as above.\footnote{Let $CE_i(t)$ be the cumulative emission data in scenario $i$ at time $t$ and $\widehat{CE}_i(t)$ be the corresponding cumulative emissions obtained by $\theta_i$. The average absolute distance is defined as $\frac{1}{n}\sum_{t=1}^{n} \vert 1- \frac{\widehat{CE}_i(t)}{CE_i(t)}\vert $. } 

\subsection{Interpretation of the decarbonization speed}\label{subsec_interpretation}
Given each decarbonization speed $\theta_i$, we can also compute the required time (in years) to halve the initial carbon intensity in each scenario.\footnote{The formular is $\ln(1-0.5/u_{\max})/-\theta_i$.} The time to halve the initial carbon intensity carries the same information as $\theta_i$ but has a more transparent interpretation. Table \ref{tab:transition} shows the scenario specific decarbonization speed $\theta_i$ and its ambition defined as years to halve the initial carbon intensity across the five selected IAM scenarios. A larger value of $\theta_i$ implies a shorter time to halve the initial carbon intensity and therefore corresponds to a more aggressive mitigation pathway. The decarbonization speed metric thus enables a ranking of IAM scenarios according to their implicit mitigation ambition. Applying the above procedure to each IAM scenario yields a total of 126 decarbonization speed estimates. In the next section, we give an overview of these estimates.
\begin{table}[!htbp]
    \centering
    \begin{tabular}{rllllll}
    \hline
    \hline
     scenarios    & SSP1-19&SSP2-34&SSP3-60&SSP4-45 &SSP5-60\\
     \hline
     fitted $\hat{\theta_i}$&0.0124&0.0842&0.0368&0.0480&0.0549\\ 
     ambition(years): &22 &37 &91&73 &63 \\
         \hline
         \hline
    \end{tabular}
    \caption{This table provides the fitted transition ambition metric $\hat{\theta_i}$ in all scenarios. We also provide the corresponding necessary time (in years) to reduce 50\% of the initial carbon intensity (recorded as ambition (years)). }
    \label{tab:transition}
\end{table}
\subsection{An overview of the decarbonization speed across IAM scenarios}
\textcolor{black}{Following the procedure described above, we estimate a unique decarbonization speed for each IAM scenario. We further compute each scenario's transition ambition, defined as the number of years to halve the initial carbon intensity implied by the estimated decarbonization speed. As some scenarios lack data for 2005, we use 2010 as the common starting point for the analysis. The initial carbon intensity is around 0.24 KgCO$_2$/kwh. At this stage, more than 85\% of primary energy supply is derived from fossil fuels. When it is halved to 0.12 KgCO$_2$/kwh, the percentage of fossil fuel-based primary energy is projected to decline to around 60\%.}

\textcolor{black}{Table \ref{tab:overview_ambition} gives an overview of the 126 estimates of the decarbonization speeds. We classify the IAM scenarios according to their transition ambition. The first column in Table \ref{tab:overview_ambition} reports the time interval to halve the initial carbon intensity and the second column gives the number of scenarios falling within each interval. The SSPs column lists the corresponding SSP assumptions in each group, where ``all'' means all SSPs appear at least once. Similarly, the RCPs column records the RCP assumptions in each group. The last column reports the IAM coverage.}

\textcolor{blue}{\begin{table}[!htbp]
    \centering
    \begin{tabular}{lllll}
    \hline
    \hline
         Ambition (years)&\# scenarios &SSPs & RCPs& IAMs  \\
         $<10$ years& 0 & n.a. & n.a.& n.a.\\
         $(10,20]$ years&31&all&RCP19/26/34 & all\\
         $(20,30]$ years &28 &all &RCP26/34/45 &all \\
         $(30,40]$ years &16 &all &RCP34/45/60 & all \\
         $(40,50]$ years &12 &all &RCP45/60 & all\\
         $(50,90]$ years &26 &all&RCP60/Baseline &all\\
         $>90$ years &13 &SSP2/3/4/5 &Baseline &all\\
         \hline
         \hline
    \end{tabular}
    \caption{An overview of decarbonization ambition across 126 scenarios.}
    \label{tab:overview_ambition}
\end{table}}

\begin{remark}
We give a few remarks on Table \ref{tab:overview_ambition}.
    \begin{enumerate}
        \item Year 2010 is the starting year. 113 out of 126 scenarios indicate that the carbon intensity can be halved by 2100. Only 13 scenarios need a longer time to achieve this goal. All scenarios require at least 10 years to halve the initial carbon intensity.
        \item Except for the highest-ambition group, all SSPs appear at least once within each ambition interval. This suggests that SSP assumptions alone do not uniquely determine the pace of the energy transition.
        \item In contrast, RCPs display a clear monotonic pattern across ambition groups. Higher RCP levels, which correspond to warmer scenarios and weaker mitigation targets, appear exclusively in the less ambitious groups, whereas lower RCP levels are concentrated in the more ambitious groups. This systematic ordering indicates that the decarbonization speed is consistent with the underlying forcing targets of IAM scenarios and provides a meaningful summary of their mitigation ambition.
        \item Within each ambition group, all IAMs are represented at least once, indicating that the observed patterns are not driven by any single modeling framework. 
    \end{enumerate}
\end{remark}

\begin{figure}[!htbp]
       \centering
       \includegraphics[width=0.75\linewidth]{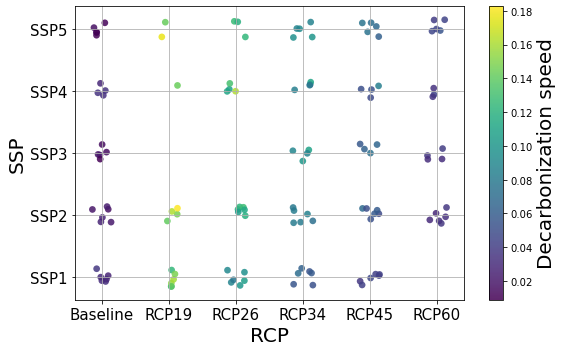}
       \caption{The scatter plot of the 126 decarbonization speed estimates.}
       \label{fig:scatter}
   \end{figure}

Figure~\ref{fig:scatter} displays a scatter plot of the corresponding decarbonization speed estimates across scenarios. For completeness, the decarbonization speed estimates for all 126 IAM scenarios are reported in the Supplementary Material. In Figure~\ref{fig:scatter}, each scenario is positioned at the intersection of its SSP and RCP categories. The colored points represent the scenario-specific decarbonization speed estimates $\theta_i$, with darker colors indicating smaller values and lighter colors indicating larger values. Multiple points may appear within the same SSP–RCP cell because different IAM modeling teams can produce alternative quantitative realizations of the same scenario narrative. For readability, these estimates are displayed with slight offsets around their corresponding SSP–RCP locations. Figure~\ref{fig:scatter} shows that scenarios associated with higher RCP levels tend to exhibit lower decarbonization speeds, indicating that the ordering of scenarios by decarbonization speed is consistent with their underlying mitigation narratives. 

\section{Stylized Distribution and Uncertainty of Decarbonization Speeds}\label{sec_parametric}
The previous section demonstrates that the decarbonization speed is a useful metric to represent the overall mitigation ambition of a climate scenario. When comparing mitigation ambition, the narrative-based high dimensional datasets are reduced to numbers that can be compared and ordered on a common scale. In this section, we give more statistical analysis on these decarbonization speed estimates.

We construct an empirical distribution of decarbonization speed and quantify uncertainty using bootstrap resampling. Since the true distribution is unknown, the bootstrap offers a convenient nonparametric framework for characterizing uncertainty in ensemble-based estimates.
\subsection{Empirical distribution of decarbonization speeds}\label{subsec_empirical}
\begin{figure}[!htbp]
       \centering
       \includegraphics[width=0.75\linewidth]{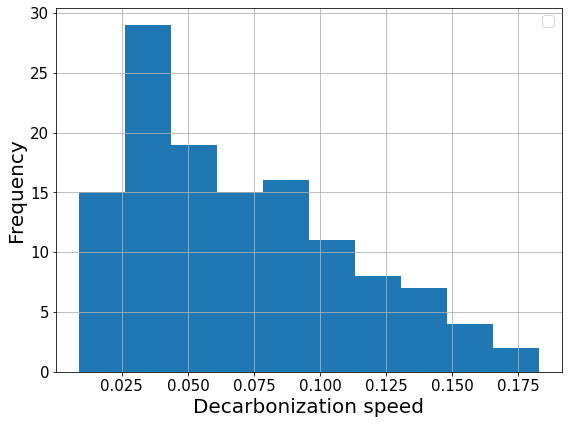}
       \caption{The histogram of the 126 decarbonization speed estimates with 10 bins. The minimum $\theta_i$ is 0.009, which needs around 225 years to halve the initial carbon intensity. The maximum $\theta_i$ is 0.18, which needs around 11 years to halve the initial carbon intensity.}
       \label{fig:Histspeed}
   \end{figure}

\textcolor{black}{In this section, we present the empirical distribution of decarbonization speed obtained from 126 IAM scenarios. Figure~\ref{fig:Histspeed} displays the histogram of the 126 decarbonization speed estimates, while Table~\ref{tab:summary} reports key summary statistics. The empirical distribution (Figure~\ref{fig:Histspeed}) is right-skewed with a long tail, indicating that most IAM scenarios imply gradual or moderate decarbonization, whereas only a small subset of scenarios correspond to very rapid decarbonization associated with highly ambitious mitigation pathways.}

\textcolor{black}{Table~\ref{tab:summary} further shows that the mean decarbonization speed is slightly larger than the median, consistent with the right-skewed shape of the distribution. Half of the IAM scenarios require more than 33 years to reduce their initial carbon intensity (0.24~kgCO$_2$/kWh at 2010) by 50\%. By comparison, historical data (Figure~\ref{fig:CarbonIntensityHistory}) indicate that the European Union, the fastest historical decarbonizing region, required approximately 60 years to halve its carbon intensity relative to its 1965 level. Among the IAM scenarios, about 70\% imply a decarbonization speed faster than the historical experience of the European Union. This comparison suggests that the IAM ensemble embodies an overall optimistic vision of future decarbonization relative to observed historical trends.}

   \begin{table}[!htbp]
       \centering
       \begin{tabular}{r|c}
       \hline
       \hline
        Mean    &  0.07 (28.5 years)\\
         Median   & 0.06 (33.1 years) \\
         Standard deviation & 0.04 (9.5 years)\\
         Min & 0.009 ( 225 years) \\
         Max & 0.18 (10.9 years) \\
         25th percentile & 0.035 (56.9 years) \\
         75th percentile & 0.099 (20.2 years) \\
            \hline
            \hline
       \end{tabular}
       \caption{Summary statistics of decarbonization speed (necessary years to halve the initial carbon intensity) across 126 IAM scenarios.}
       \label{tab:summary}
   \end{table}
   This section presents the empirical distribution of the 126 decarbonization speed estimates together with key summary statistics. Because these estimates are obtained from a finite ensemble of scenarios, they are subject to uncertainty. The next section introduces a bootstrap resampling technique to quantify the uncertainty associated with these ensemble-based estimates. 
\subsection{Bootstrap confidence band}\label{subsec_bootstrap}
\textcolor{black}{The previous section reports the decarbonization speed estimates for the 126 IAM scenarios, including their empirical distribution and key summary statistics such as the mean, median, and selected quantiles. As emphasized by Mudelsee (2019) \cite{mudelsee2019trend}, statistical estimates without accompanying uncertainty quantification are of limited interpretative value. We therefore apply bootstrap resampling to construct confidence intervals for these summary statistics. Because the bootstrap does not rely on assumptions about the functional form of the underlying true distribution, it is well suited to our setting.} 

\textcolor{black}{Bootstrap is a nonparametric resampling technique, which has wide applications in many fields. One common application is to approximate the sampling distribution of an estimator (such as mean and median) when the true underlying distribution is unknown. By repeatedly drawing samples with replacement from the observed data and recomputing the statistic of interest, the bootstrap method provides a practical means of estimating uncertainty and constructing confidence intervals without imposing strong distributional assumptions. For technical properties and examples of the bootstrap method, we recommend the book by Efron and Tibshirani \cite{efron1994introduction}.}

\textcolor{black}{In the field of climate science, the bootstrap method has been widely used to quantify the uncertainty in the ensemble and simulation study. For example, O'Reilly et al (2020)  \cite{o2020calibrating} use historical data to calibrate the large-ensemble European climate projections. In particular, the bootstrap method is used to quantify the uncertainty when examining the performance of calibrated ensembles on out-of-sample data. Basso et al (2025) \cite{basso2025multi} use bootstrap to quantify the uncertainty of the multi-model ensemble median prediction in soil-carbon storage efficiency. Uhe et al (2021) \cite{uhe2021method} compare different modeling frameworks for precipitation projections and use the bootstrap to quantify the uncertainty in mean and extreme precipitation projections. Smith et al (2015) \cite{smith2015probabilistic} analyze a European Multimodel Ensemble system for seasonal climate forecast and use the bootstrap to build confidence interval for the ensemble mean. We also recommend \cite{portmann2025climloco1} and \cite{candille2007verification} for the application of the bootstrap to quantify the ensemble uncertainty.\footnote{In addition to ensemble uncertainty quantification, the bootstrap method is also widely used for quantifying uncertainty in climate time series data. For this type of application, we recommend \cite{mudelsee2019trend}, \cite{varga2017generalised}, \cite{innocenti2022analytical} and \cite{paciorek2018quantifying}. Moreover, the bootstrap resampling can also be used for improving out-of-sample prediction. For this, we recommend \cite{taillardat2016calibrated}. }} 

We take the median estimate as an example and demonstrate how to use the bootstrap method to quantify its uncertainty. Table \ref{tab:summary} reports that the median of the 126 decarbonization speed estimates is 0.06, or equivalently, it means that half of the IAM scenarios need more than 33 years to halve the initial carbon intensity level. Since the underlying distribution of the decarbonization speed is unknown, there is no closed-form formula to compute the uncertainty of the median estimate. 

\begin{figure}[!htbp]
       \centering
       \includegraphics[width=0.75\linewidth]{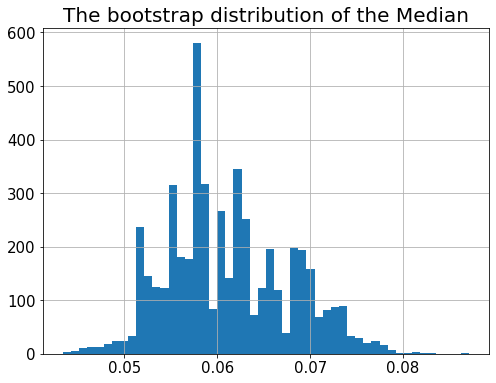}
       \caption{The histogram of the median generated by 5000 bootstrap samples. }
       \label{fig:MedianBoot}
   \end{figure}

The 126 decarbonization speed estimates can be represented in the following way:
\begin{equation}
    \{(S1,\theta_1), (S2,\theta_2),...,(S126, \theta_{126})\}.
\end{equation}
Since each IAM scenario is uniquely determined by its SSP, RCP and modeling team, we give a label to each scenario, $S1,S2,\cdots,S126$. Then, we randomly sample from the 126 scenario labels with replacement and formulate a new sample with 126 scenarios, which is called a bootstrap sample. Note that the same scenario can appear more than once. Figure \ref{fig:MedianBoot} provides a histogram of the median estimates by 5000 bootstrap samples.\footnote{According to Efron and Tibshirani\cite{efron1994introduction}, a few thousand bootstrap samples are typically sufficient in most cases to obtain a distribution of the statistic.} The mean of the median estimates by the bootstrap samples is 0.061, which is very close to the original median estimate. The 5\% and 95\% percentiles of the median estimates are 0.052 and 0.073,  corresponding to approximately 38 and 27 years to halve the initial carbon intensity, respectively. We conclude that the median estimate is reasonably stable with respect to the ensemble variations, with a bootstrap interval $(0.052,0.073)$, or equivalently, the median time to halve the initial carbon intensity falls in $(27 \text{years}, 38 \text{years})$.

Following the same procedure, we provide in Table \ref{tab:summaryCI} the bootstrap confidence intervals for some other summary statistics. For interpretability, in the lower panel, we translate the decarbonization speed metric to the required time to halve the initial carbon intensity. We remark that a larger value of the decarbonization speed implies a more ambitious mitigation schedule and therefore a shorter time horizon to halve the initial carbon intensity. For example, the expected time to halve the initial carbon intensity is about 28.5 years. Considering the ensemble uncertainty, the expected time may be between 26.2 years and 31.3 years.  Overall the bootstrap intervals for these estimated statistics are reasonably tight, suggesting that these estimates are stable against ensemble variations. 

    \begin{table}[!htbp]
       \centering
       \begin{tabular}{r|c|c}
       \hline
       \hline
       statistic & original ensemble estimate & bootstrap-CI (5\%,95\%)\\
       \hline
        Mean    &  0.07 & (0.064, 0.076)\\
         Median   & 0.06 &(0.052, 0.073) \\
         25th percentile & 0.035 & (0.032, 0.041) \\
         75th percentile & 0.099 & (0.087, 0.110) \\
         \hline
         Mean    &  28.5 years & (26.2 years, 31.3 years)\\
         Median   & 33.1 years &(27.4 years, 38.4 years) \\
         25th percentile & 56.9 years & (49.2 years, 61.6 years) \\
         75th percentile & 20.2 years & (18.1 years, 23.0 years) \\
            \hline
            \hline
       \end{tabular}
       \caption{Upper panel: Representative statistics of the decarbonization speed with the confidence interval computed from bootstrap distribution. Lower panel: interpretation as the time to halve the initial carbon intensity, i.e., $\ln(1-\frac{0.5}{u_{\max}})/(-\theta_i)$.}
       \label{tab:summaryCI}
   \end{table}
\subsection{Fitting a stylized parametric distribution to decarbonization speeds}\label{subsec_fit}
\textcolor{black}{In this section, we fit a parametric distribution to the 126 estimates of the scenario decarbonization speed. This stylized parameteric representation provides a high level summary of the transition patterns outlined by IAM scenarios. In this way, complicated scenario narratives and high-dimensional datasets can be reduced to only a few parameters. Representing IAM scenarios by a parametric distribution will also simplify comparison and quantitative analysis of IAM scenarios. Ultimately, it may facilitate applications to fields beyond climate science. However, a parameteric distribution is usually concise and cannot capture all aspects of climate scenarios. We will discuss these limitations in the next section.}

\textcolor{black}{By definition, the decarbonization speed is non negative, i.e., $\theta\geq 0$. In addition, its empirical distribution (Fig. \ref{fig:Histspeed}) is slightly right skewed. Therefore, we consider the candidate parametric distributions that have a long right tail and a non negative support. Candidate distributions are for example, Lognormal, Gamma,  and Weibull etc.}

\textcolor{black}{Many parametric functions defined on the positive real line can be used to approximate the empirical distribution. However, our goal is not to identify the best fitting model. Fitting multiple parametric distributions would introduce additional technical details. Instead, our goal is to demonstrate how the IAM scenarios can be summarized by a simple stylized parametric function. For this purpose, we only use a lognormal distribution as an illustration.}

\textcolor{black}{\begin{definition}{Lognormal distribution}\\
    Let $X$ be a non negative random variable defined on the positive real line. It follows a lognormal distribution with parameters $\mu$ and $s$, 
    \begin{equation}
        X \sim LN(\mu, s^2) \quad \text{or equivalently},\quad \ln{X}\sim N(\mu, s^2).
    \end{equation}
    Its probability density function is given by 
    \begin{equation}
        f(x) = \frac{1}{xs\sqrt{2\pi}}\exp\left(-\frac{(\ln x-\mu)^2}{2s^2}\right).
    \end{equation}
    Its mean and variance are given by
    \begin{equation}
        \mathbb E[X]=\exp\left(\mu+\frac{s^2}{2}\right),\quad Var(X) = (\exp(s^2)-1)\exp(2\mu+s^2).
    \end{equation}
\end{definition}
A lognormal distribution is characterized by two parameters $\mu$ and $s$. Suppose the decarbonization speed follows a lognormal distribution, i.e., $\theta_i\sim LN(\mu,s^2), i=1, 2, \cdots 126$. Thus, the Maximum-Likelihood-Estimates (MLE) of the pamameters of the lognormal distribution are given by
\begin{equation}
    \hat{\mu}=\frac{1}{126}\sum_{i=1}^{126}\ln\theta_i,\quad \widehat{s^2}=\frac{1}{126}\sum_{i=1}^{126}(\ln\theta_i-\hat{\mu})^2.
\end{equation}
Plugging in the decarbonization speed estimates $\theta_i$ (provided in the Supplementary Material), we have the following:
\begin{equation}
    \hat{\mu} = -2.87\quad\text{and}\quad \widehat{s^2} = 0.48.
\end{equation}}

\begin{figure}[!htbp]
       \centering
       \includegraphics[width=0.75\linewidth]{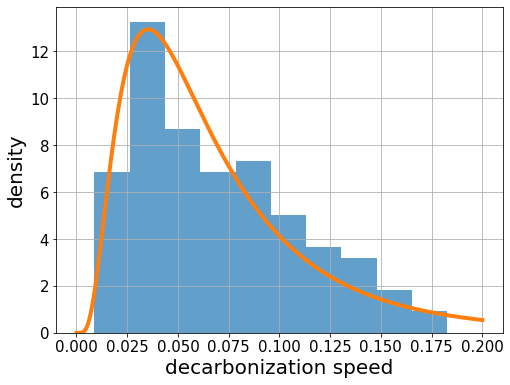}
       \caption{The fitted lognormal density function. }
       \label{fig:lognormaldensity}
   \end{figure}

\begin{figure}[!htbp]
       \centering
       \includegraphics[width=0.75\linewidth]{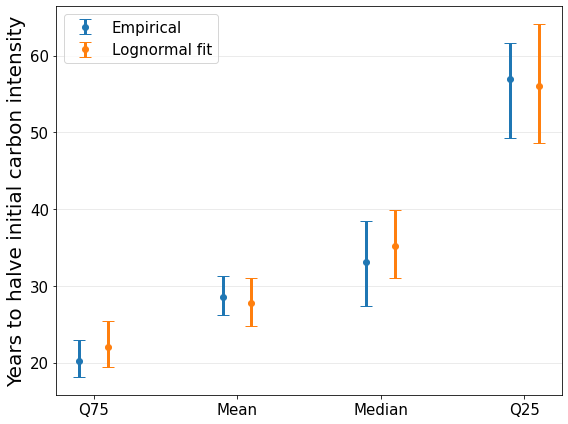}
       \caption{Comparison of the empirical estimates and the fitted lognormal distribution. The error bars are given by the 5th and 95th value from the 5000 bootstrap resamples. For the empirical estimates, we use a nonparametric bootstrap while for the fitted lognormal distribution, we use the parametric bootstrap. }
       \label{fig:logfit}
   \end{figure}

\textcolor{black}{Figure \ref{fig:lognormaldensity} illustrates the fitted lognormal density and the empirical distribution. The fitted curve captures well the main features of the decarbonization speed distribution. Moreover, Table \ref{tab:logfit} reports the key summary statistics such as mean and median computed from the fitted lognormal distribution. We also use the bootstrap resampling to derive the confidence interval of these statistics. Different from the previous section, we use a parametric bootstrap resample technique. Assuming that the decarbonization speed follows a lognormal distribution, we randomly sample 126 decarbonization speeds from the fitted model and create a bootstrap sample. Then we compute the corresponding statistics such as mean and median. Repeating this 5000 times, we obtain 5000 values of the key statistics, of which the 5th and 95th values are used to create the confidence intervals. Fig. \ref{fig:logfit} compares the key statistics from the empirical estimates and the fitted lognormal distribution. The fitted lognormal distribution matches well with the empirical distribution.}  

    \begin{table}[!htbp]
       \centering
       \begin{tabular}{r|c|c}
       \hline
       \hline
       statistic & fitted lognormal & bootstrap-CI (5\%,95\%)\\
       \hline
        Mean    &  0.072 & (0.065, 0.079)\\
         Median   & 0.057 &(0.051, 0.063) \\
         25th percentile & 0.036 & (0.031, 0.041) \\
         75th percentile & 0.090 & (0.082, 0.099) \\
         \hline
         Mean    &  27.7 years & (25.3 years, 30.5 years)\\
         Median   & 35.2 years &(31.8 years, 38.9 years) \\
         25th percentile & 56.0 years & (49.1 years, 63.4 years) \\
         75th percentile & 22.1 years & (20.2 years, 24.4 years) \\
            \hline
            \hline
       \end{tabular}
       \caption{This table reports the statistics computed from the fitted lognormal distribution.}
       \label{tab:logfit}
   \end{table}
\subsection{Limitations of the current 
approach}\label{subsec_limits}
\textcolor{black}{The main contribution of this work is the introduction of a simple numerical metric of the decarbonization speed of an IAM scenario. With this metric, the narrative based IAM scenarios and the high-dimensional time series datasets can be ranked and compared on a common scale. The previous section demonstrates that these estimates can be represented by a stylized parametric distribution, which may facilitate quantitative analysis of the IAM scenarios by fields beyond climate science. In this section, we discuss some caveats and limitations of this approach.
\begin{enumerate}
    \item The decarbonization speed is defined to measure the improvement of carbon intensity, which mainly concerns the primary energy and the CO$_2$ emissions data. The main motivation for this choice is that many other variables can be viewed as either upstream drivers of energy data or downstream consequences of emissions data. Nevertheless, this metric leaves other scenario variables unpresented.\\ 
    \item The scenario variables are reported as time series data. The carbon intensity we computed are also time series. However, the decarbonization metric is rather a smooth measure of the overall trajectory. Considering the fact that the scenario variables are projected at coarse time intervals on a ten-year step size, such a smooth measure captures well decarbonization trends in most scenarios. However, by design, it cannot capture abrupt shifts in the decarbonization trends. Abrupt policy adjustments are indeed a feature in some scenario framework such as NGFS, which cannot be represented well with the current definition of the decarbonization speed.\\
    \item To represent the 126 IAM scenarios concisely, we construct the empirical distribution and a fitted parametric distribution. The purpose of these stylized parametrization is to enable more advanced quantitative analysis of the climate scenarios, which represents the view on possible carbon intensity improvements. When making probabilistic assessment for the potential transition pathway, the key driver may go beyond carbon intensity improvement. Therefore, this stylized distribution should be used with caution. 
\end{enumerate}}
\section{Conclusion}
This work studies the 126 IAM scenario datasets modeled by six leading teams in climate science. Our main contribution is to define a simple numerical metric that measures the decarbonization speed of each IAM scenario. Through this metric, the high-dimensional time series scenario datasets can be compared and ranked in a transparent way. We further construct the empirical distribution and a fitted parametric distribution of the decarbonization speed estimates. In addition, the key statistics such as mean and median of the estimates with bootstrap confidence intervals are provided. The stylized distribution of IAM scenarios focuses on the view on the carbon intensity improvement, which is a central dimension in mitigation. The simple parameterization may also facilitate further quantitative analysis of the climate scenarios.
\newpage
\bibliography{up}

\end{document}